\documentclass[preprintnumbers,prd,twocolumn,floatfix,preprintnumbers,letterpaper,amsmath,amssymb,nofootinbib,superscriptaddress]{revtex4}
\usepackage{graphicx}
\usepackage{epsfig}
\usepackage{bm}
\usepackage{amsfonts}
\usepackage{color}
\usepackage{subfigure}

\usepackage{amsmath,amssymb}

\begin{document}

\title{Prospect of probing dark energy using stochastic gravitational waves}

\author{Bikash R Dinda}
\email{bikash@ctp-jamia.res.in}
\affiliation{Centre for Theoretical Physics, Jamia Millia Islamia, New Delhi-110025, India}

\author{Anjan A Sen}
\email{aasen@jmi.ac.in}
\affiliation{Centre for Theoretical Physics, Jamia Millia Islamia, New Delhi-110025, India}

\begin{abstract}
We study the possibility of probing dark energy behaviour using gravitational wave experiments like LISA and Advanced LIGO. Using two popular parameterizations for dark energy equation of state, we show that with current sensitivities of LISA and Advanced LIGO to detect the stochastic gravitational waves, it is possible to probe a large section of parameter space for the dark energy equation of state which is allowed by present cosmological observations. 
\end{abstract}

\maketitle
\date{\today}

\maketitle

\section{Introduction}

Notwithstanding the evidence of late time acceleration of the Universe from a variety of cosmological observations \cite{SN, CMB, BAO}, the lack of any definitive answer so far for the cause of this acceleration is quite troubling. That is why the goal for most of the current and future observational efforts is to unravel this mystery. The idea is to study the effect of late time acceleration on different cosmological observables like distances, angles, volumes, growth of structures and many more. Subsequently measuring these observables with great accuracies, one can constrain the different dark energy \cite{DE} or modified gravity models \cite{MG} (the two most widely studied approach to explain the late time acceleration). In this approach, unless one accurately handles the systematics involved in different cosmological observations, it is very difficult to get any meaningful constraint on dark energy models. After the latest results by Planck in 2015 \cite{Planck15} combined with other non-cmb measurements \cite{SN,BAO}, we now have the most stringent constraint on dark energy behaviour. This shows that the latest cmb+non-cmb data is not only consistent with simple cosmological constant($\Lambda$), it is also consistent with other non-$\Lambda$ behaviours and more interestingly with phantom behaviour \cite{Hazra, Planck15}. As the issue is not settled yet, we are looking for data from future observational set ups like EUCLID \cite{Euclid}, DES \cite{DES}, SKA \cite{SKA} etc, which may hopefully single out a particular class of dark energy behaviour decisively.

Under this circumstances, it is always useful to look for other new observables that can be useful probes for dark energy. In this work, we study the prospects of probing dark energy models by observing the stochastic gravitational wave background. Gravitational waves provides a unique opportunity to see our Universe in a way that is not possible through cosmological observations \cite{RGW1}. In recent times, lot of progress has been made in the process of detecting gravitational waves using laser interferometer. The Laser Interferometer Space Antenna (LISA) \cite{LISA} and the Advanced Laser Interferometer Gravitational Waves Observatory (Advanced LIGO) \cite{ALIGO} will provide a major breakthrough in gravitational wave astronomy. The chance to detect gravitational wave including stochastic gravitational wave has increased manyfold using these detectors. Hence it is interesting to study the effect of dark energy on the stochastic gravity wave spectrum and subsequently the possibility to probe dark energy models using the data from these two detectors (See \cite{RGW2, RGW3} for earlier studies in this direction).

In this work, using two popular parameterizations for dark energy equation of state, we study the possibility of probing dark energy behaviour using LISA and Advanced LIGO. We show that with the current  
sensitivities of these two detectors, it will be possible to probe a large portion of parameter space that is currently allowed by the present cmb+non-cmb data. 

Taking into account the tensorial perturbation in a flat Friedman-Robertson-Walker metric given by

\begin{equation}
 ds^{2} = a^{2}(\tau)[d\tau^{2}-(\delta_{ij}+h_{ij})dx^{i} dx^{j}] ,
 \end{equation}

\noindent
where $h_{ij}$ is the tensorial gravitational field and is transverse and traceless. For a fixed wave vector $k$ and a fixed polarization, $h_{ij}$ is governed by the wave equation \cite{RGW1, RGW2, RGW3}

\begin{equation}
\mu_{k}^{''} + (k^{2}-\frac{a^{''}}{a}) \mu_{k} = 0
\end{equation}

\noindent
where $ h_{k}(\tau) = \mu_{k}(\tau)/a(\tau) $ and prime denotes derivative w.r.t the conformal time $\tau$. Subsequently the spectrum of the relic gravitational waves $ h(k,\tau) $ can be calculated using $ h(k,\tau) = \frac{4 l_{Pl}}{\sqrt{\pi}} k |h_{k}(\tau)| $, where $ l_{Pl} $ is the Planck length. Once $ h(k,\tau) $ is known, energy density $ \Omega_{g}(k) $ of the relic gravitational waves is given by \cite{RGW1, RGW2, RGW3}

\begin{equation}
\Omega_{g}(k) = \frac{\pi^{2}}{3} h^{2}(k,\tau_{0}) \Big{(}\frac{k}{k_{0}} \Big{)}^{2}.
\end{equation}

Here $ k_{0} $ is the wave number corresponding to the present Hubble radius. To solve, equation (2), one has to assume a initial contribution for the relic gravitational wave background from inflation. The current data from temperature anisotropy in the CMB measured by Planck gives an upper limit to $r$ which is the ratio of the amplitudes of the tensor and scalar fluctuations produced during inflation. Assuming the observed temperature anisotropy in CMB $\Delta T/T$ to be $\approx 0.37\times 10^{-5}$ at multipole $l=2$ \cite{Planck_inflation} (corresponding to anisotropies on Hubble scale), the spectrum $h(k,\tau)$ is fixed at present as

\begin{equation}
h(k_{0},\tau_{0}) = 0.37 \times 10^{-5} \times r,
\end{equation}

\noindent
where $\tau_{0}$ is the conformal time at present. We assume $r=0.1$ in our calculations which is the upper limit on $r$ as observed by Planck \cite{Planck_inflation}.

Next we need to know the scale factors $a(\tau)$ at different epochs in the Universe to solve equation (2). For the first four stages of the universe, e.g inflation (I), reheating (RH), radiation dominated (R) and matter dominated (M), we assume the scale factors as \cite{RGW2,RGW3}:

\begin{eqnarray}
&& a_{I} (\tau)=l_{i} |\tau -\tau_{i}|^{1+\alpha}, \hspace{0.8 cm} -\infty < \tau \leq \tau_{I}, \nonumber\\
&& a_{RH} (\tau)=a_{z} (\tau -\tau_{p})^{1+\beta}, \hspace{0.4 cm} \tau_{I} \leq \tau \leq \tau_{RH}, \nonumber\\
&& a_{R} (\tau)=a_{e} (\tau -\tau_{e}), \hspace{1.2 cm} \tau_{RH} \leq \tau \leq \tau_{R}, \nonumber\\
&& a_{M} (\tau)=a_{m} (\tau -\tau_{m})^{2}, \hspace{0.7 cm} \tau_{R} \leq \tau \leq \tau_{M},
\end{eqnarray}

\noindent
where $\alpha$ determines the inflationary evolution. For $\alpha=-2$, the inflationary period is exactly de-Sitter. But as we know that the inflationary period is not exact de-Sitter but a small deviation from that, we assume $\alpha=-1.95$ for our subsequent calculations. This value is consistent with the previous work by Zhang et al \cite{RGW2, RGW3} on spectrum of gravitational waves assuming $\Lambda$CDM model. $\beta$ determines the expansion during the reheating stage and $\beta \neq -1$ \cite{RGW1}. The parameters $\tau_{I}, \tau_{RH}, \tau_{R}$ and $\tau_{M}$ determine the different epochs. 

To get the scale factor during the late time accelerating epoch, we solve the Friedmann equation,

\begin{equation}
H^{2} = H_{0}^{2} \Big{[} \Omega_{m}^{(0)} a^{-3} + \Omega_{DE}^{(0)} \exp{\Big{\{}-3(\int_{a_{0}}^{a} da^{'} \frac{1+w(a^{'})}{a^{'}})\Big{\}}}\Big{]}, 
\end{equation}

\noindent
for time interval $ \tau_{M} \leq \tau \leq \tau_{0} $.  $ H_{0} $, $ \Omega_{m}^{(0)} $ $ \& $ $ \Omega_{DE}^{(0)} $ are the present day Hubble parameter and density parameters for matter and dark energy respectively. We assume a spatially flat Universe which is consistent with recent Planck result \cite{CMB}. $w(a)$ is the dark energy equation of state and in this work we consider two most popular parameterizations for $w(a)$. These are the CPL \cite{CPL} parameterization and the GCG parameterization \cite{GCG} given respectively by 

\begin{eqnarray}
w(a) &=& w_{0} + w_{a} (1-a),\\
w_{g}(a) &=& - \frac{A_{s}}{A_{s}+(1-A_{s})a^{-3(1+\alpha)}},
\end{eqnarray}

\noindent
where $w_{0}, w_{a}, A_{s}$ and $\alpha$ are the parameters involved in these parameterizations. One can show that the GCG parameterization incorporates two interesting quintessence behaviour e.g. the thawing ($1+\alpha <0$) and tracking behaviour ($1+\alpha >0$) \cite{GCG}

All the constants in eq. (5) will be fixed with respect to junction times by taking normalization $ a(\tau_{0}) = a_{0} = 1 $ and then by equating $ a $ and $ da/d\tau $ at the junction times. Moreover, to fix the junction times, we assume $ a(\tau_{0})/a(\tau_{M}) = 1.33 $, $ a(\tau_{M})/a(\tau_{R}) = 3454 $, $ a(\tau_{R})/a(\tau_{RH}) = 10^{24} $, $ a(\tau_{RH})/a(\tau_{I}) = 300 $ \cite{RGW2,RGW3}.

\begin{figure*}[t]
\begin{tabular}{c@{\quad}c}
\epsfig{file=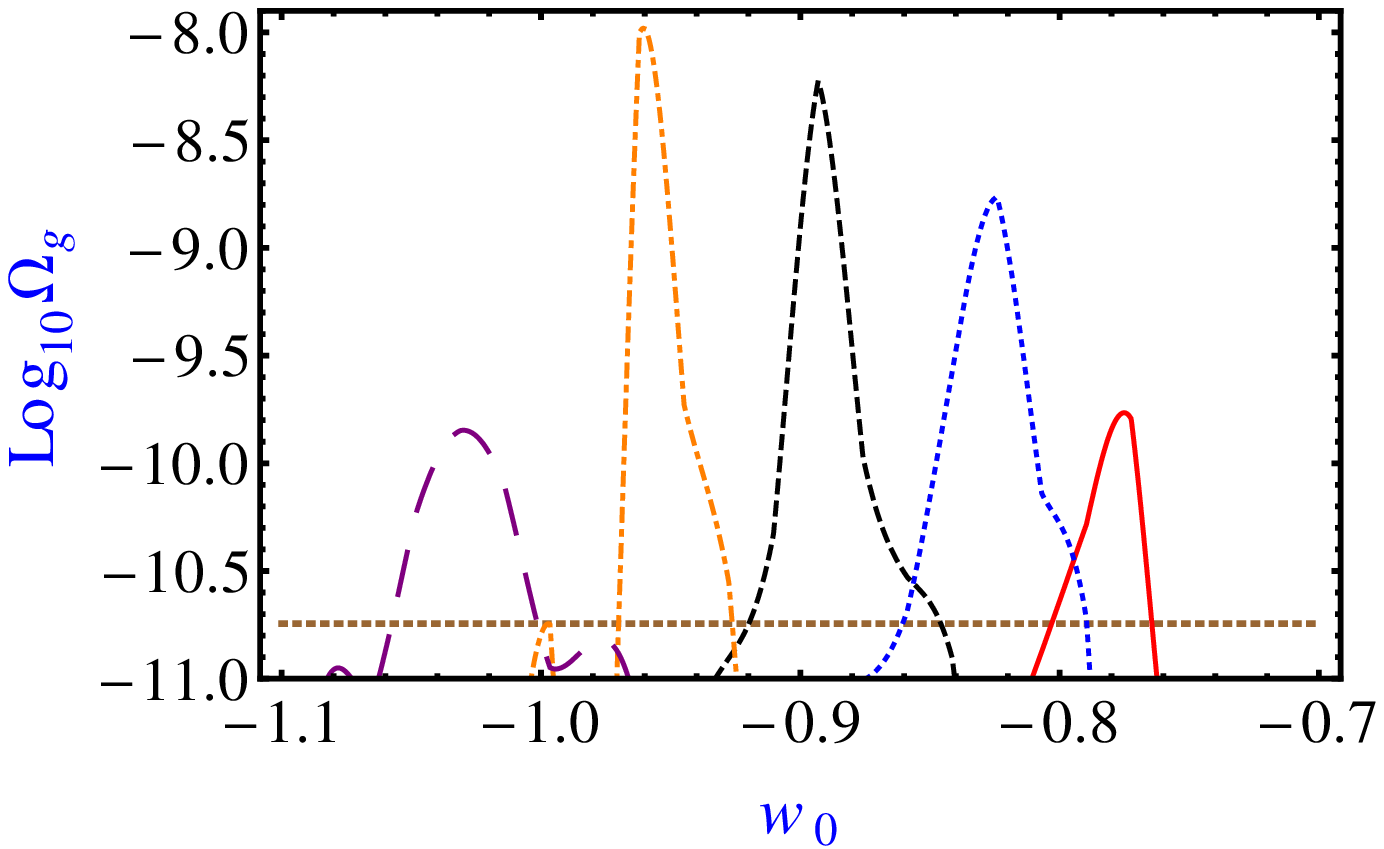,width=7.5 cm,height=5 cm}
\epsfig{file=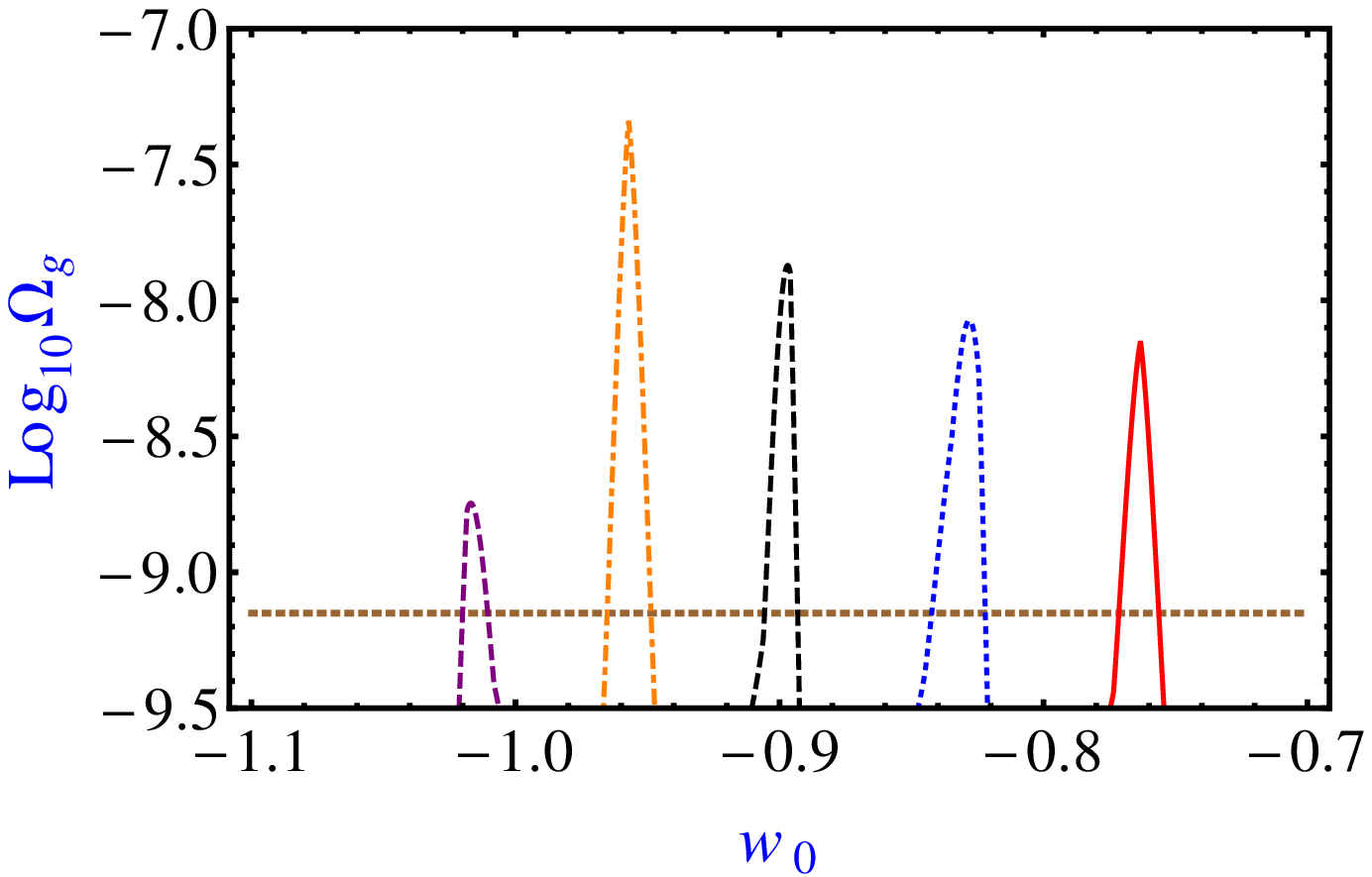,width=7.5 cm, height=5 cm}
\end{tabular}
\caption{CPL Parameterization: $ \Omega_{g} $ vs. $ w_{0} $ graphs: Left and right panels are at the frequencies $ 2.92\times 10^{-3} $ Hz (i.e. lowest point of the LISA sensitivity curve) and 64.56 Hz (i.e. lowest point of the advanced LIGO sensitivity curve) respectively. For both the graphs continuous-red, dotted-blue, dashed-black, dot-dashed-orange and long-dashed-purple lines are for $ w_{a} $ values -1, -0.5, 0, 0.5 and 1 respectively. Brown-dotted lines in left and right panels are for the amplitude of $ \Omega_{g} $ at lowest point of the LISA and advanced LIGO sensitivity curves respectively.
}
\end{figure*}


\begin{center}
\begin{figure*}[t]
\begin{tabular}{c@{\quad}c}
\epsfig{file=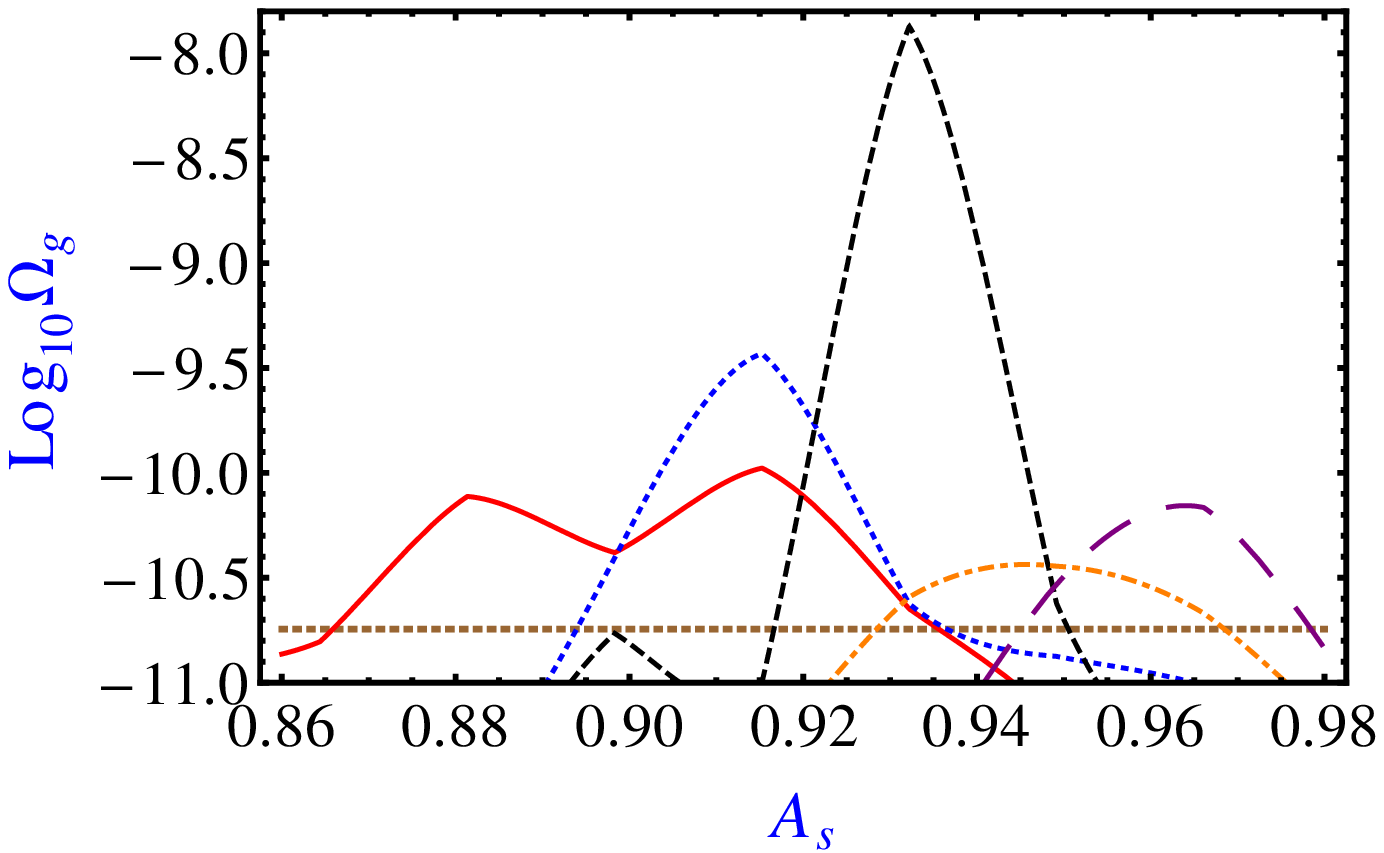,width=7.5 cm,height=5 cm}
\epsfig{file=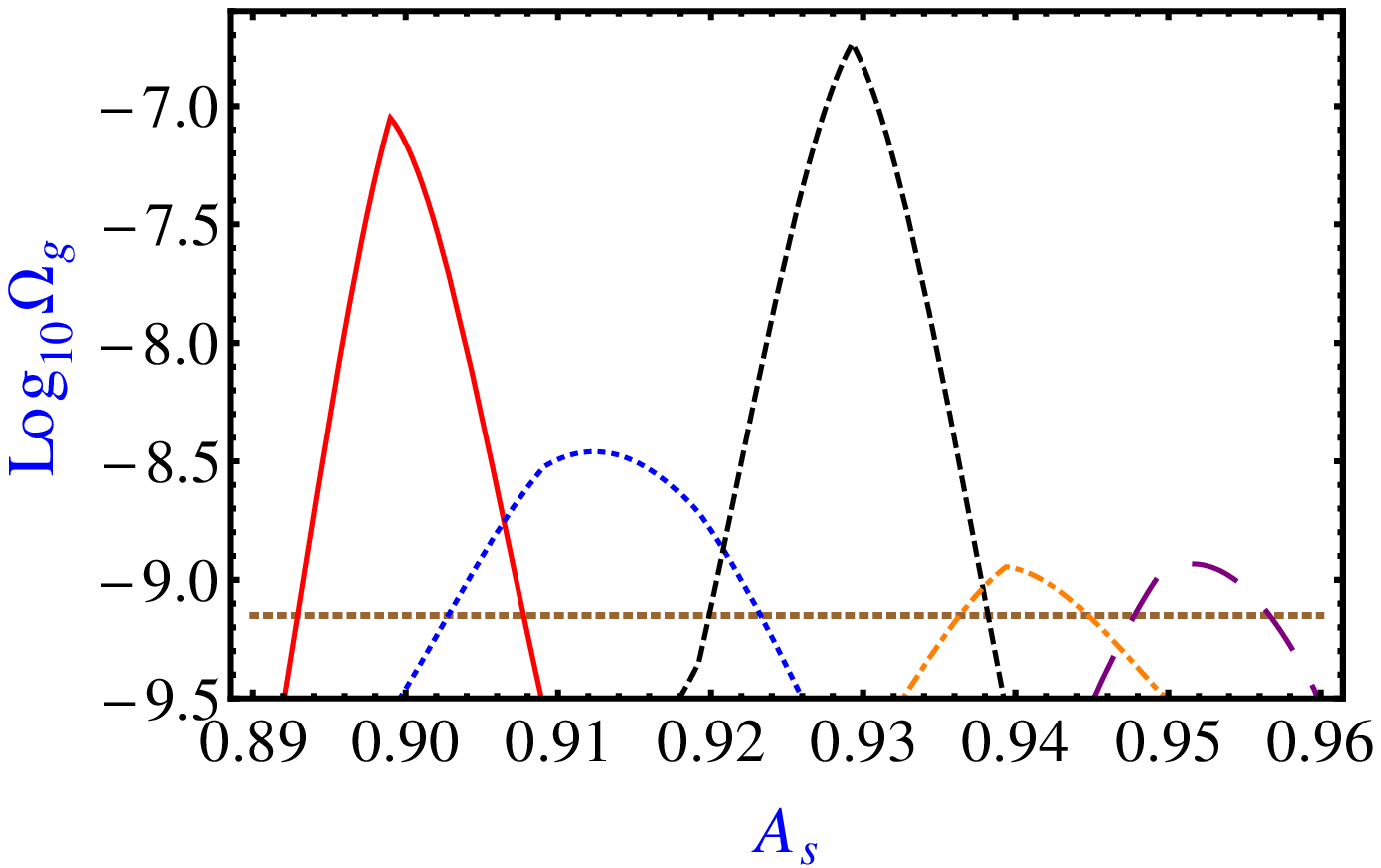,width=7.5 cm,height=5 cm}
\end{tabular}
\caption{GCG Parameterization: $ \Omega_{g} $ vs. $ A_{s} $ graphs: Left and right panels are at the frequencies $ 2.92\times 10^{-3} $ Hz (i.e. lowest point of the LISA sensitivity curve) and 64.56 Hz (i.e. lowest point of the advanced LIGO sensitivity curve) respectively. For both the graphs continuous-red, dotted-blue, dashed-black, dot-dashed-orange and long-dashed-purple lines are for $ \alpha $ values -0.9, -0.5, 0, 0.5 and 0.9 respectively. Brown-dotted lines in left and right panels are for the amplitude of $ \Omega_{g} $ at lowest point of the LISA and advanced LIGO sensitivity curves respectively.
}
\end{figure*}
\end{center}


\section{Solution of the spectrum}

For scale factors of power law form e.g. $ a(\tau) \propto \tau^{\alpha} $, the solution of eq. (2) has the form

\begin{equation}
\mu_{k}(\tau) = \sqrt{k\tau} \Big{[}c_{1} J_{\alpha-1/2}(k\tau)+c_{2} J_{1/2-\alpha}(k\tau)\Big{]},
\end{equation}

\noindent
where $ c_{1} $ and $ c_{2} $ are two arbitrary constants.

So, the inflationary, reheating, radiation-dominated and matter-dominated stages have the solutions respectively as

\begin{eqnarray}
&& \mu_{k}^{I}(\tau) = \sqrt{x_{1}} \Big{[}A_{1} J_{\beta+1/2}(x_{1})+A_{2} J_{-(\beta+1/2)}(x_{1})\Big{]}, \nonumber\\
&& \mu_{k}^{RH}(\tau) = \sqrt{x_{2}} \Big{[}B_{1} J_{\beta_{s}+1/2}(x_{2})+B_{2} J_{-(\beta_{s}+1/2)}(x_{2})\Big{]}, \nonumber\\
&& \mu_{k}^{R}(\tau) = \sqrt{x_{3}} \Big{[}C_{1} J_{1/2}(x_{3})+C_{2} J_{-1/2}(x_{3})\Big{]}, \nonumber\\
&& \mu_{k}^{M}(\tau) = \sqrt{x_{4}} \Big{[}D_{1} J_{3/2}(x_{4})+D_{2} J_{-3/2}(x_{4})\Big{]},
\end{eqnarray}

\noindent
where $ x_{1} = k |\tau-\tau_{i}| $, $ x_{2} = k(\tau-\tau_{p}) $, $ x_{3} = k(\tau-\tau_{e}) $ and $ x_{4} = k(\tau-\tau_{m}) $.\\

\noindent
For Inflationary stage, we can use the adiabatic initial condition, $ \lim_{k \rightarrow \infty} \mu_{k}(\tau) \rightarrow e^{- i k \tau} $ which gives $ A_{1} = -\frac{i}{\cos{\beta \pi}} \sqrt{\frac{\pi}{2}} e^{i \pi \beta/2} $ and $ A_{2} = i A_{1} e^{- i \pi \beta} $. The arbitrary constants $ B_{1} $, $ B_{2} $, $ C_{1} $, $ C_{2} $, $ D_{1} $ and $ D_{2} $ can be fixed by equating $ \mu_{k}(\tau) $ and $ \frac{d \mu_{k}}{d\tau} (\tau) $ at the junction times.  Moreover $\beta$ can be fixed using the normalization condition given by eqn (4). After determining all the coefficients we get $ \mu $ and $ d\mu/d\tau $ at $ \tau_{M} $ which are given by 

\begin{eqnarray}
\mu_{k}^{Late} \Big{|}_{ini} &&= - \sqrt{\frac{2}{\pi}} \frac{1}{k z_{1}}\Big{\lbrace} \Big{[} k z_{1} \cos(k z_{1}) - \sin(k z_{1}) \Big{]} D_{1} \nonumber\\
&&+ \Big{[} \cos(k z_{1}) + k z_{1} \sin(k z_{1}) \Big{]} D_{2} \Big{\rbrace}, \nonumber\\
\frac{d \mu_{k}^{Late}}{d \tau}\Big{|}_{ini} &&= \sqrt{\frac{2}{\pi}} \frac{1}{k z_{1}^{2}} \Big{\lbrace}\Big{[} k z_{1} D_{1} + (1-k^{2} z_{1}^{2}) D_{2} \Big{]} \cos(k z_{1}) \nonumber\\
&& + \Big{[}(k^{2} z_{1}^{2} -1) D_{1} +   k z_{1} D_{2} \Big{]} \sin(k z_{1}) \Big{\rbrace},
\end{eqnarray}

\noindent
where $ z_{1} = (\tau_{M} - \tau_{m}) $. Using these initial conditions we will solve eq. (2) numerically during the epoch $\tau_{M} \leq \tau \leq \tau_{0}$ using the Hubble parameter during the dark energy dominated epoch given by eqn (6).

\section{Results}

We fix $ \Omega_{DE}^{(0)} = 0.7 $, $ \Omega_{m}^{(0)} = 0.3 $, $ \alpha =-1.95 $ and $ r = 0.1 $ in our calculations. These values are consistent with the latest Planck results \cite{CMB}.

In fig. (1) we show $\Omega_{g} $ as a function of $ w_{0}$  for different values of $w_{a}$ for CPL parameterization. It depends on the frequency of the observed gravitational waves. In the left panel, we plot it for $\nu = 2.92\times 10^{-3}$ Hz which is the frequency corresponding to the lowest bound for LISA sensitivity curve \cite{LISA_sens}. The horizontal line corresponds to the lowest bound for $\Omega_{GW}$ for LISA. In the right panel, we do the same for Advanced LIGO with the corresponding frequency fixed at $\nu = 64.56$ Hz \cite{LIGO_sens}. The regions above the horizontal line are detectable for the present sensitivity for LISA and Advanced LIGO. In other words, if LISA and Advanced LIGO measures gravitational waves with these sensitivity at the corresponding frequencies, then the regions above the horizontal line can be probed and constrained by gravity wave experiments. In fig. (2), we redo the analysis for GCG parameterization where we plot $\Omega_{g} $ as function of $ A_{s}$ for different values of $\alpha$. We should point out that for GCG, $-A_{s}$ represents the present value $w_{0}$ of the dark energy equation of state.  

\begin{center}
\begin{figure*}[t]
\begin{tabular}{c@{\quad}c}
\epsfig{file=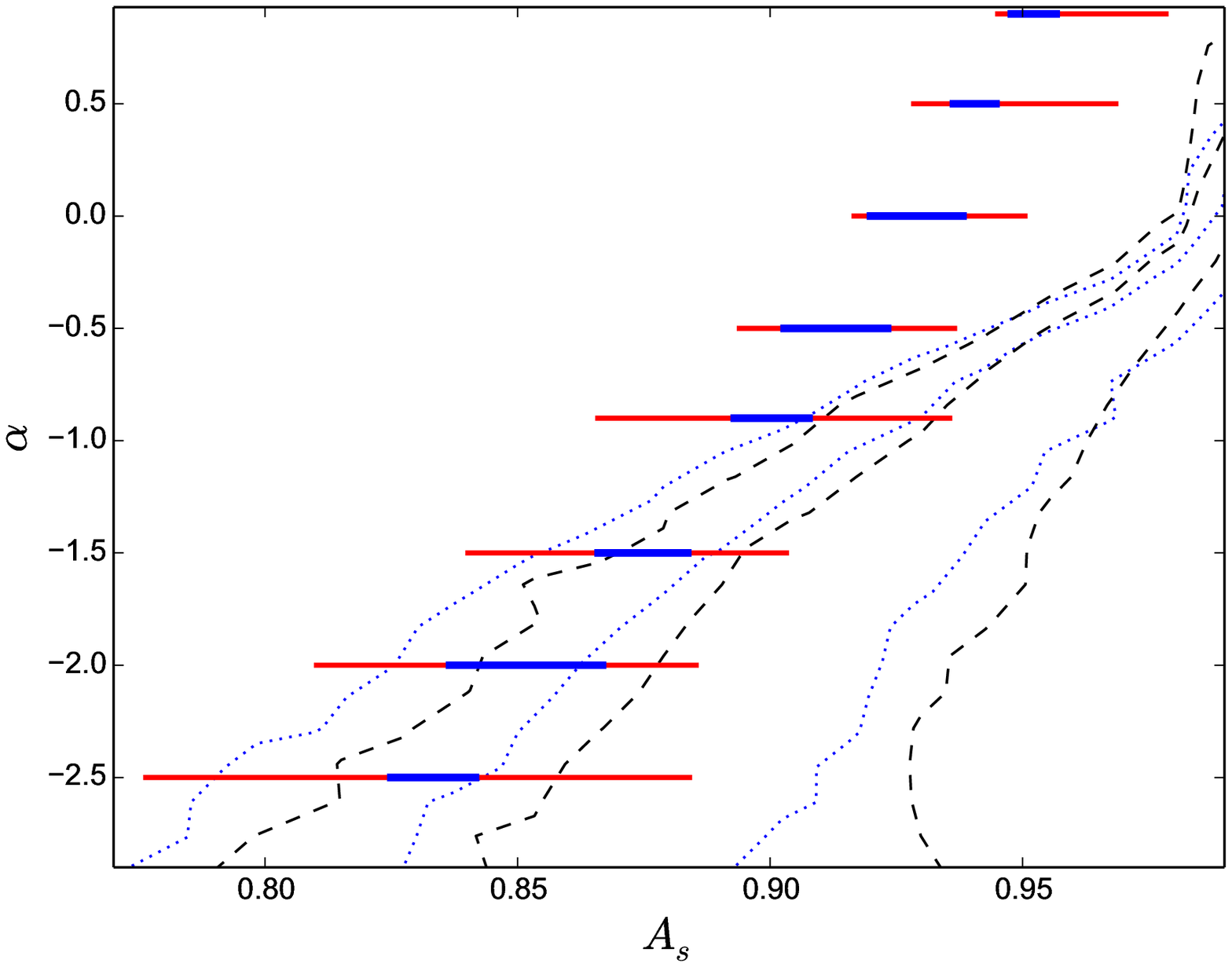,width=7.5 cm,height=5 cm}
\epsfig{file=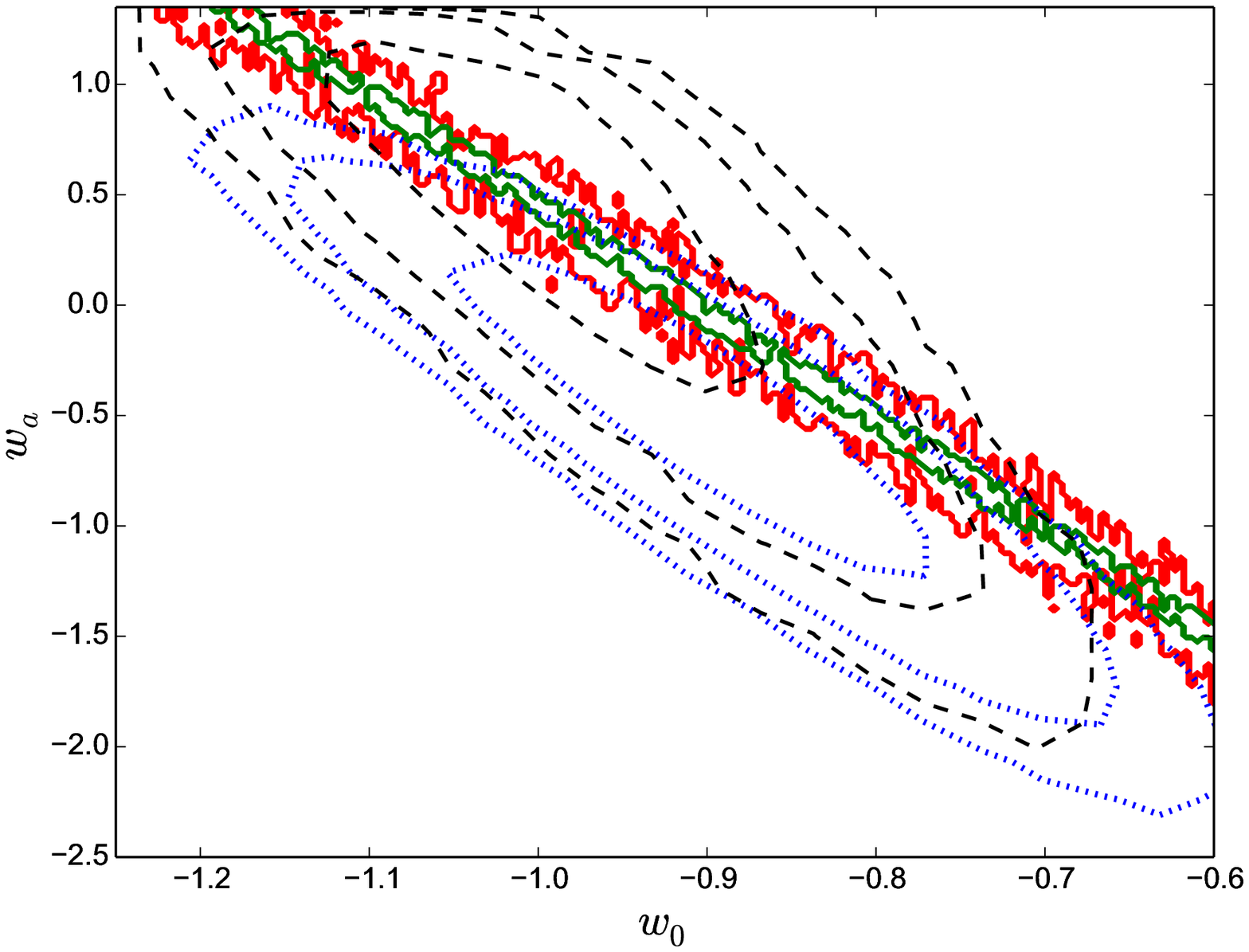,width=7.5 cm,height=5 cm}
\end{tabular}
\caption{Left Panel: The allowed parameter space for GCG parameterization for JLA(SN)+BAO+CMB-Shift (dashed) and JLA(SN)+BAO+CMB-Shift+$f\sigma_{8}$ (dotted) data (See \cite{Sumit} for details about the data and the corresponding analysis). The Horizontal lines are those probed by LISA(bigger lines) and Advanced LIGO(smaller lines). Right Panel:The allowed parameter space for CPL parameterization for JLA(SN)+BAO+CMB-Shift (dashed) and JLA(SN)+BAO+CMB-Shift+$f\sigma_{8}$ (dotted) data. The Solid outer(inner) regions are those that can be probed by LISA(Advanced LIGO) experiments. 
}
\end{figure*}
\end{center}

Next we show how these regions occur in the current allowed parameter space. In fig. (3), we show this for GCG and CPL parameterizations. In the left panel, we show the current allowed region in the $A_{s}-\alpha$ parameter space for GCG for the present observational data  from SN+BAO+CMB-Shift+$f\sigma_{8}$ (for details about the data and the subsequent analysis, please see \cite{Sumit}). In the same plot, we show the regions from fig.(2), that can be probed by LISA and Advanced LIGO in future. It is clear that for GCG, future gravity wave experiments can only probe thawing ($1+\alpha< 0$) region. Most of the tracker region ($1+\alpha > 0$) allowed by the cosmological data is away from the region that can be probed by gravity waves except a very narrow region near $\alpha \sim -1$. For CPL parameterization, we show the same in the right panel. Here the regions that can be probed by LISA or Advanced LIGO fall right inside the region that is currently allowed by the cosmological data. We should stress that the CPL parameterization is used by almost all the cosmological observations to put constraints on dark energy. Hence the fact that the future gravity wave experiments can probe a considerable region in its parameter space that is allowed by present cosmological data is extremely encouraging.

Although we fix $r=0.1$ in our calculations which is the upper bound on $r$ by Planck, but the results do not differ much even if we have lower values for $r$ e.g $r=0.05$. Moreover we fix $\beta=-1.95$ in our calculations which fixes the cosmological evolution during inflation. For smaller values of $\beta$, the regions that can be probed by gravity wave experiments become smaller and for larger values of $\beta$, the regions get bigger.

\section{Conclusion}

To conclude, we show the possibility of constraining dark energy equation of state using the future gravitational waves experiments. Using two popular parameterizations for dark energy equation of state, we show that detectors like LISA and Advanced LIGO can have the power to probe regions in the parameter space which are currently allowed by different cosmological observations. For GCG case, mostly the thawing region can be probed by the experiments whereas for the CPL parameterization which is the universal parameterization for dark energy equation of state used by all cosmological observations, future gravity wave experiments can probe a large section of the parameter space that is currently allowed by the cosmological observations. Our investigation is not fully robust as we fix different cosmological parameters to do the analysis. But probably for the first time, our results show the promise of probing relevant parameter space for dark energy behaviour using gravity wave experiments. This definitely motivates one to do more detailed analysis in this regard.
\vspace{3mm}
\section{Acknowledgements}
The authors would like to thank Md. Wali Hossain for extensive discussions. The authors also thank Sumit Kumar for providing the plots for the constraint regions for CPL and GCG using cosmological data. We also acknowledge the use of publicly available sensitivity data for advanced LIGO and LISA. B.R.D. thanks CSIR, Govt. of India for financial support through SRF scheme (No:09/466(0157)/2012-EMR-I).

\end{document}